\documentclass[graybox]{svmult}

\usepackage{type1cm}        % activate if the above 3 fonts are
\usepackage{makeidx}         % allows index generation
\usepackage{graphicx}        % standard LaTeX graphics tool
\usepackage{multicol}        % used for the two-column index
\usepackage[bottom]{footmisc}% places footnotes at page bottom

\usepackage{newtxtext}       % 
\usepackage{newtxmath}       % selects Times Roman as basic font

\makeindex             % used for the subject index
\begin{document}

\title*{The challenge of scale in molecular adaptation: Local searches in astronomical genotype networks}
\titlerunning{The challenge of scale in molecular adaptation}
%\author{Susanna Manrubia \and Luis F. Seoane \and J. A. Cuesta}
%\institute{Susanna Manrubia \at Museo Nacional de Ciencias Naturales (CSIC), c/ Jos\'e Gutiérrez Abascal 2, 28006 Madrid, Spain, \email{smanrubia@mncn.csic.es}
%\and Luis F. Seoane \at Institut de Biologia Evolutiva (UPF-CSIC), Pg. Mar\'{\i}tim de la Barceloneta 37, 08003 Barcelona, Spain, \email{lf.seoane@ibe.upf-csic.es}
%\and Jos\'e A. Cuesta \at Universidad Carlos III de Madrid, Departamento de Matemáticas, Grupo Interdisciplinar de Sistemas Complejos (GISC), Avda. de la Universidad 30, 28911 Legan\'es, Spain, \email{cuesta@math.uc3m.es}}
\author{
Susanna Manrubia\inst{1,4} \and
Luis F. Seoane\inst{2,4} \and
Jos\'e A. Cuesta\inst{3,4}
}
\institute{
$^1$Museo Nacional de Ciencias Naturales (CSIC), 
c/ Jos\'e Gutiérrez Abascal 2, 28006 Madrid, Spain\\
\email{smanrubia@mncn.csic.es} \\
$^2$Institut de Biologia Evolutiva (UPF-CSIC), 
Pg. Mar\'{\i}tim de la Barceloneta 37, 08003 Barcelona, Spain\\
\email{lf.seoane@ibe.upf-csic.es} \\
$^3$Universidad Carlos III de Madrid, Departamento de Matemáticas, 
Avda. de la Universidad 30, 28911 Legan\'es, Spain\\
\email{cuesta@math.uc3m.es} \\
$^4$Grupo Interdisciplinar de Sistemas Complejos (GISC), Madrid, Spain
}

%\and Name of Second Author \at Name, Address of Institute \email{name@email.address}}
\maketitle

\abstract*{The exploration of vast genotype spaces poses fundamental challenges for evolving populations. As the number of genotypes encoding viable phenotypes grows exponentially with genome length, populations can only explore a tiny fraction of these immense spaces, a fact consistently supported by empirical and theoretical evidence. Paradoxically, local, mutation-driven searches near abundant sequences allow populations to generate phenotypic improvements and functional innovations despite this immense search space. In this contribution, we integrate insights from viral evolution with theoretical expectations derived from genotype-phenotype maps to re-examine how high-dimensional sequence spaces shape evolutionary dynamics. In resolving the paradox, abundant phenotypes play a crucial role because their combinatorial weight biases evolutionary trajectories. We discuss how this bias, together with limited accessibility of fitness peaks, modifies traditional metaphors---such as fitness landscapes--- and challenges standard notions of evolutionary optimality. Our results underscore that adaptation is predominantly local yet remarkably efficient, providing a unifying perspective on the coexistence of robustness, innovation, and constrained exploration in molecular evolution.}

\abstract{
%The exploration of vast genotype spaces poses fundamental challenges for evolving populations. Although the number of genotypes encoding viable phenotypes grows exponentially with genome length, empirical and theoretical evidence shows that populations explore only a tiny fraction of these spaces. In this contribution, we integrate insights from viral and microbial populations with models of genotype-to-phenotype maps and neutral networks, emphasizing how high-dimensional sequence spaces shape evolutionary dynamics. Local, mutation-driven searches near abundant sequences allow populations to generate phenotypic improvements and functional innovations despite the immense theoretical space. We highlight the role of abundant phenotypes, whose combinatorial weight biases evolutionary trajectories and discuss how this effect, together with limited peak accessibility, modifies traditional metaphors such as fitness landscapes and notions of evolutionary optimality. Our results underscore that adaptation is predominantly local yet remarkably efficient, providing a unifying perspective on the coexistence of robustness, innovation, and constrained exploration in molecular evolution.\\
The exploration of vast genotype spaces poses fundamental challenges for evolving populations. As the number of genotypes encoding viable phenotypes grows exponentially with genome length, populations can only explore a tiny fraction of these immense spaces, a fact consistently supported by empirical and theoretical evidence. Paradoxically, local, mutation-driven searches near abundant sequences allow populations to generate phenotypic improvements and functional innovations despite this immense search space. In this contribution, we integrate insights from viral evolution with theoretical expectations derived from genotype-phenotype maps to re-examine how high-dimensional sequence spaces shape evolutionary dynamics. In resolving the paradox, abundant phenotypes play a crucial role because their combinatorial weight biases evolutionary trajectories. We discuss how this bias, together with limited accessibility of fitness peaks, modifies traditional metaphors---such as fitness landscapes--- and challenges standard notions of evolutionary optimality. Our results underscore that adaptation is predominantly local yet remarkably efficient, providing a unifying perspective on the coexistence of robustness, innovation, and constrained exploration in molecular evolution.}

\section{Introduction}
\label{sec:1}

Genetic diversity is a fundamental prerequisite for species survival and evolution \cite{agashe:2023}. Natural populations produce and maintain high levels of diversity through multiple mechanisms, ensuring efficient exploration of functional alternatives and rapid responses to changing environments. Our knowledge of intra-species diversity is particularly extensive for viruses and microbial populations. High-throughput sequencing and improved sampling have repeatedly revealed substantial genetic diversity within single hosts \cite{lauring:2020}, within microbial species in complex environments, and across replicate experimental populations. Deep sequencing of SARS-CoV-2 \cite{jones:2025} and other RNA viruses \cite{borderia:2011} has identified measurable intra-host single-nucleotide variants \cite{seoane:2025}, persistent mixed infections in some patients, and lineage-specific differences in within-host diversity \cite{gu:2023}. These studies and many others indicate that viral populations commonly exist as diverse ``clouds'' of related genotypes consistent with a quasispecies-like structure \cite{domingo:2006}. 

Despite the typically lower mutation rates of prokaryotes and smaller effective population sizes compared with viral populations, metagenomics and genome-resolved sequencing have revealed high standing diversity within bacterial species in natural microbiomes \cite{kim:2024}. Many species are not represented by a single dominant genotype but by coexisting strain lineages and within-strain polymorphisms \cite{viver:2024}. Long-read and single-cell approaches uncover fine-scale structures—haplotypes, mobile elements, and recombination tracts—that substantially increase inferred within-taxon diversity relative to earlier short-read-based snapshots. These data indicate that bacteria {\it in situ} maintain far more genetic variation than simple single-clone models predicted \cite{harris:2021}. 

{\it In vivo} observations are complemented by {\it in vitro} evolution experiments, which show that, in large asexual populations, adaptation often involves multiple coexisting beneficial lineages \cite{lang:2013}, polygenic shifts, and frequent soft sweeps \cite{harris:2021}. Populations experience both rapid selective sweeps and extended periods of balanced polymorphism or competing mutations \cite{laguna-castro:2025}, depending on population size, mutation supply, recombination, and ecological structure. These findings suggest that periods away from mutation–selection equilibrium may be the natural state of microbial \cite{cvijovic:2018} and viral \cite{braun:2018} populations. 

Several theoretical frameworks have been developed to explain these observations. Kimura's Neutral Theory \cite{kimura:1968} proposed that most molecular substitutions are effectively neutral, with substitution rates equal to mutation rates—a powerful null model for sequence divergence. Ohta extended this to the nearly-neutral regime \cite{ohta:1973}, emphasizing that the fate of weakly selected mutations depends on population size and drift. Maynard Smith's protein space metaphor \cite{maynard-smith:1970} and subsequent work on neutral networks \cite{schuster:1994} showed that many genotypes can have similar fitness and that connectivity via neutral mutations provides a mechanism to perform a sustained exploration of sequence space without fitness losses. In parallel, Eigen introduced the quasispecies framework \cite{eigen:1971,eigen:1989}, describing high-mutation-rate populations of replicators \cite{schuster:1983,schuster:1988}, later applied to RNA viruses as clouds of related genotypes surrounding a master sequence \cite{domingo:1978}. In quasispecies populations, the effects of the underlying fitness landscape are attenuated by collective dynamics, allowing coexistence of mutants with diverse fitness values both in and out of mutation–selection equilibrium. 

Efficient generation of genetic diversity is essential for long-term adaptability. Mechanisms that maintain extant diversity—such as high mutation rates or transposition bursts in response to unexpected selection pressures \cite{piacentini:2014}—have likely been selected over evolutionary timescales \cite{ferrare:2024,kumawat:2025}. In light of these observations, several key questions arise: To what extent are neutral spaces essential for evolvability? How constrained are populations to occupy fitness maxima? Are ``valleys'' in fitness landscapes true barriers to adaptation in high-dimensional, networked genotype spaces? And to what extent does finite population size limit the discovery of functional phenotypes?

\section{Neutral networks and genotype-to-phenotype maps}
\label{sec:2}

\subsection{Synthetic genotype-to-phenotype maps}

Maynard Smith's idea of protein networks has been highly influential in subsequent characterizations and descriptions of both the structure of sequence spaces (induced by a variety of genotype-to-phenotype (GP) maps \cite{schuster:1994,lipman:1991,arias:2014,greenbury:2014,fortuna:2017}) and the supposed constraints operating along evolutionary dynamics. In all analyses of GP maps there is the underlying assumption that mutations causing significant variation in the expressed phenotype will be rare or absent in populations, since they are deleterious and therefore selected against. Two ideas should be kept in mind in this context. First, natural selection acts on phenotypes, so significant evolutionary changes will only occur when phenotypes change. Since the fitness value of a phenotype is context-dependent, it is generally not possible to measure the value of a phenotype without testing its performance in specific environments. More often than not, even in constant environments, non-neutral mutations whose adaptive value is difficult to quantify are fixed in populations \cite{villanueva:2022}, occasionally modifying phenotypes in unforeseeable ways \cite{escarmis:2008,grant:2021}. Secondly, the presence of deleterious mutations in a population results from a competition between two temporal scales: the rate of generation of such mutations and the strength of purifying selection. This tension between the two scales is particularly relevant in ensembles of mutationally linked replicators, such as viruses, where the fast mutational scale translates into a persistent coexistence of variants bearing mutations with dissimilar effects on fitness \cite{lauring:2020} and even different phenotypes \cite{duarte:1994,vignuzzi:2006}. 

\begin{figure}
    \centering
    \includegraphics[width=0.65\linewidth]{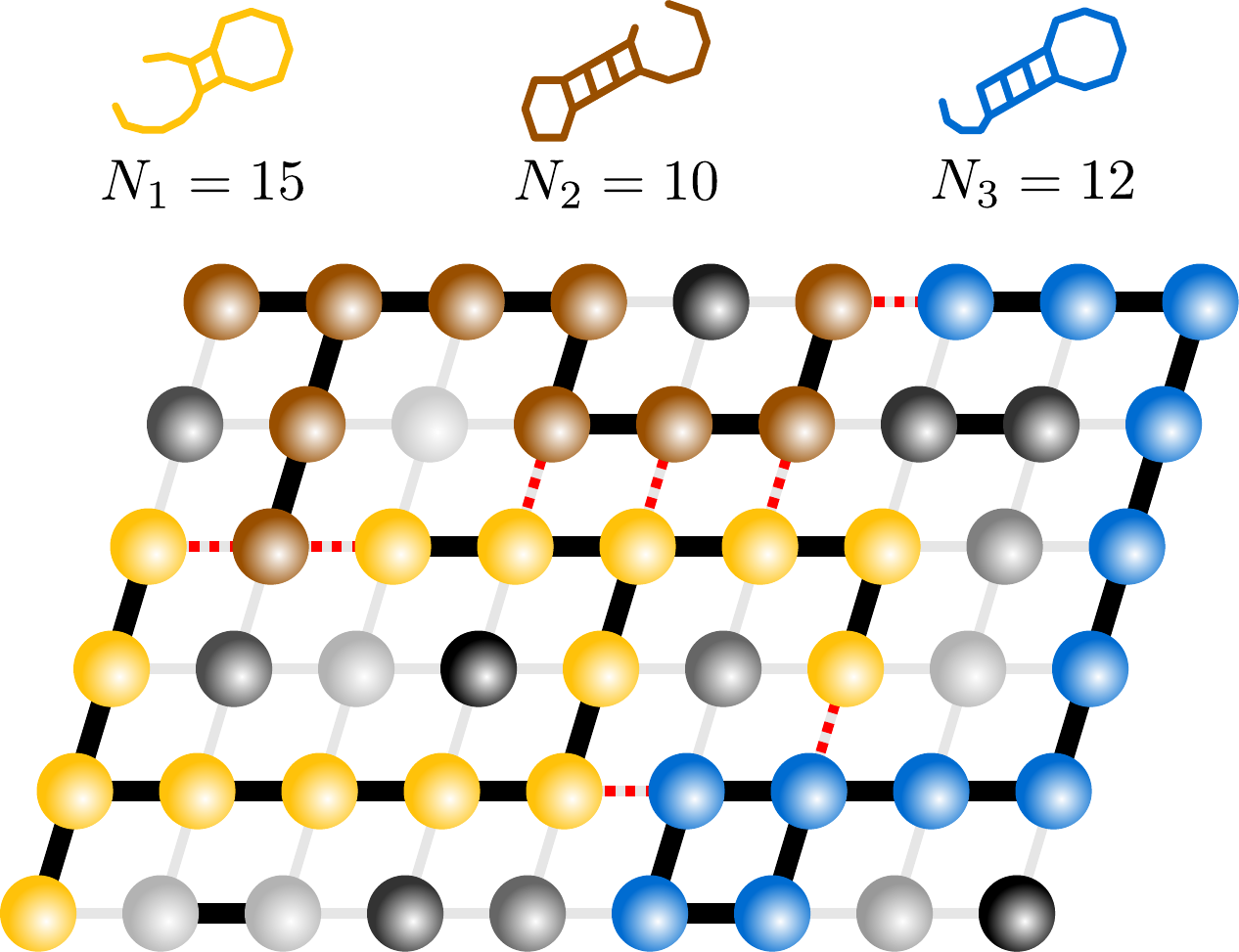}
    \caption{{\bf Neutral networks in the RNA sequence-to-secondary structure genotype-phenotype map.} 
    This grid provides a simplified representation of genotype networks, which in reality are far more complex and not planar. Each node represents a unique RNA sequence, and a single point mutation corresponds to a displacement to a neighboring node. All nodes of the same color (connected by thick black edges) represent RNA sequences that fold into the same secondary structure---i.e., different genotypes mapping to the same phenotype—and together constitute that phenotype's {\em neutral network}. Three phenotypes (yellow, brown, and blue) occupy most of genotype space. The size of a neutral network ($N_1 = 15$ for the yellow folding, $N_2 = 10$ for the brown folding, and $N_3 = 12$ for the blue folding) is the number of sequences that produce the same phenotype. Many other, much smaller neutral networks (shown in various shades of gray) correspond to much rarer foldings, produced by only a few sequences (at most one or two in this example). Mutations within a neutral network allow divergence without losing a successful phenotype (hence without risking a fitness loss). Neutral networks usually traverse genotype space (e.g.\ the blue folding spans from top to bottom) and they get in touch with many other large phenotypes at one point or another (dashed red edges).}
\label{fig:RNA}
\end{figure}

The previous observations notwithstanding, simple GP models have been tremendously enlightening for developing a more accurate picture of how genotypes map onto phenotypes. There are some properties shared by all navigable GP maps \cite{garcia-martin:2018}, such as a negative correlation between genotypic evolvability and genotypic robustness, a positive correlation between phenotypic evolvability and phenotypic robustness, and a linear growth of phenotypic robustness with the logarithm of the number of genotypes expressing the same phenotype (what is usually referred to as `phenotype size') \cite{manrubia:2021}. One of the most striking results of such models has been the quantification of phenotypic bias: most of sequence space maps to a few highly abundant phenotypes, while an astronomical number of phenotypes are rare (few sequences map to them) and essentially invisible to evolution. These quantities are illustrated in Fig.~\ref{fig:RNA} with the iconic example of the RNA sequence-to-secondary-structure GP map. 

Phenotypic bias---the huge difference in phenotype sizes---hinted at by numerical analyses of GP maps, can be analytically derived in a general scenario where only one feature of the GP map matters: the fact that different sites in a sequence have different degrees of versatility (or neutrality) while yielding the same phenotype \cite{manrubia:2017,garcia-martin:2018,martin:2022,catalan:2023}. The size of a phenotype can be approximated by the product of single-site versatility along the sequence, thus generating a log-normal distribution of phenotype sizes in almost any realistic GP map. Interestingly, various studies have shown that only large phenotypes (the rightmost tail of the log-normal distribution) are selected in natural systems \cite{jorg:2008,dingle:2015}. This highly relevant observation has two main consequences: the size of a phenotype must be incorporated as an important feature of its fitness, and the comparatively fewer large phenotypes are much more easily found through blind mutation than rare phenotypes---implying that highly frequent phenotypes dominate evolutionary dynamics \cite{cowperthwaite:2008,schaper:2014,catalan:2020,martin:2024}. 

Neutral networks of large phenotypes have at least two additional evolutionary advantages: they are more robust on average, since the average number of neutral changes (the average degree of the neutral network $\langle k \rangle$) accepted by a sequence in a phenotype of size $N$ is $\langle k \rangle \propto \log N$ \cite{aguirre:2011,catalan:2023}; and they also percolate the space of sequences, guaranteeing navigability and access to steadily growing novelty along evolution \cite{huynen:1996,huynen:1996b,buchholz:2017}.

Still, what is the quantitative relevance of these potential advantages? 
\subsection{Phenotypic robustness and abundance}

Let us begin by critically examining the effects of the increase in robustness with phenotype size, using the case of the RNA sequence-to-secondary structure map, where the relevant quantities have been analytically and numerically worked out. The goal of the following exercise is to estimate the fraction of single-point mutations that maintain the phenotype (thick black edges in Fig.\ \ref{fig:RNA}) and the complementary fraction (i.e. those mutations causing a change in phenotype, light gray or dashed red edges in Fig.\ \ref{fig:RNA}) for sequences of length $L$. This will quantitatively illustrate the difference between rare phenotypes (where most mutations are non-neutral, gray nodes in Fig.\ \ref{fig:RNA}) and typically large phenotypes (colored nodes in Fig.\ \ref{fig:RNA}), as found in natural systems. 

On the basis of structural properties of RNA secondary structure, the expected number of neutral neighbors in a network of $N$ genotypes mapping to the same secondary structure is, asymptotically in $N$ \cite{catalan:2023},
\begin{equation}
\langle k \rangle = c \log N + \mathcal{O}\big((\log N)^{1/2}\big) \, ,
\label{eq:k}
\end{equation}
with $c \simeq 1.88$ \cite{aguirre:2011}. Bounds to minimum and maximum values of the coefficient $c$ can be analytically obtained, yielding $c=1$ and $c=3/\log 4 \simeq 2.16$, respectively \cite{aguirre:2011}. In order to estimate the average degree of frequent phenotypes, we need to estimate their average size $N_{\text{freq}}$ as a function of sequence length. Analytical estimations show that the absolute number $M_p$ of abundant phenotypes (those accumulating a fraction $p>0.99$ of genotypes) grows as $M_p \simeq 1.33 L^{-2} (1.68)^L$ \cite{catalan:2023}. A rough estimate of the typical size of those phenotypes is therefore given by 
\begin{equation}
N_{\text{freq}}= \frac{4^L}{M_p} \simeq 0.75 L^2 (2.38)^L \, .    
\label{eq:Na}
\end{equation}
Neither $M_p$ nor $N_{\text{freq}}$ are significantly sensitive to the specific value of $p$, as long as $p \simeq 1$. Substituting the latter expression in Eq.~(\ref{eq:k}), we obtain the expected neutrality for typical frequent phenotypes,
\begin{equation}
\langle k \rangle \simeq 0.87 c L + \mathcal{O}\big(L^{1/2}\big) \simeq 1.64 L + \mathcal{O}\big(L^{1/2}\big) \, .  
\end{equation}
Now, since every sequence of length $L$ has $3L$ possible neighbors one mutation away, the expected fraction of neutral neighbors becomes $\rho_a = \langle k \rangle/(3L) \simeq 1.64 /3  = 0.55$, with a correction $\mathcal{O}\big(L^{-1/2}\big)$. Asymptotically, therefore, we expect that around $45\,\%$ of mutations yield new RNA secondary structures. 

The latter figure compares well with that inferred from numerical studies, where the typical size of large, abundant phenotypes has been calculated \cite{dingle:2015}. Using the distribution of phenotype sizes corresponding to natural non-coding RNAs \cite{kin:2007}, the numerically estimated fraction $\rho_n$ of neutral mutations (for sequence lengths $40<L<126$) reaches about $\rho_n \simeq 0.6$, so around $40\,\%$ of mutations cause changes in phenotype.\footnote{Data in Fig.~3 of \cite{dingle:2015} yields a numerica estimate of the size $N$ for each length $L$; the average neutral degree of a phenotype of size $N$ is then estimated as $\langle k \rangle_n = 1.88 \log N$, from which we get $\rho_n=\langle k \rangle_n/(3L)$; $\rho_n$ is a rough average over the values obtained for large $L$.}

Interestingly, both analytical $\rho_a$ and numerical $\rho_n$ estimations yield values independent of sequence length, revealing that robustness is not a quantity that can be tuned through insertions or deletions. In practice, the increase in neutrality of large phenotypes (with respect to small phenotypes, which are not observed in natural systems) implies that the mutation rate is slightly less effective in finding phenotypic novelty, requiring two or three attempts more than small phenotypes to escape the neutral network. It would be no surprise to learn that the mutation rate has been tuned along evolution to guarantee evolvability in large phenotypes---apparently the only ones occurring in natural populations.   

\subsection{Percolation and navigability in neutral networks}

Multiple theoretical observations indicate that the neutral networks of large phenotypes contain sequences as distant as two random sequences could be. This property, together with the high dimensionality of sequence spaces (and the concomitant intertwining of phenotypes), conceptually guarantees that most (frequent) phenotypic alternatives are a few mutations away from at least some genomic region of the current phenotype. These properties result in network apposition---there are points of contact between any pair of phenotypes \cite{fontana:2002} (dashed red edges in Fig.~\ref{fig:RNA}), a property empirically demonstrated \cite{schultes:2000}---and in the related property of shape space covering (novelty is a few mutations away) \cite{gruner:1996,gruner:1996b}. What, then, is the role played by percolating phenotypes? Do populations need to explore neutral networks extensively in order to adapt successfully? This is not to suggest that neutral drift is irrelevant—on the contrary, it plays an important role \cite{bloom:2007b,erdogan:2024}. Rather, the question concerns the extent of neutral drift required for adaptation and innovation. We illustrate this point below with a quantitative example.

Consider one of the shortest functional genomes, that of {\it Avocado sunblotch viroid} (ASBVd), with 246 nucleotides \cite{symons:1981}. The total number of different sequences of length $L=246$ is $4^L \simeq 10^{148}$. A smaller figure corresponds to the number of sequences compatible with the RNA secondary structure of the ASBVd folded genome, with about 180 paired and 66 unpaired nucleotides, which is about $5 \times 10^{41}$ \cite{garcia-martin:2018,catalan:2019a}: still, the mass of all possible sequences in the ASBVd neutral network is about $10^4$ times the total biomass on Earth. How many different sequences could have been produced since this viroid appeared? An upper bound can be estimated if we assume (i) ASBVd appeared simultaneously with land plants, some 470 My ago; (ii) the number of infected plants at any time is $10^{12}$;\footnote{Viroids infect a wide range of cultivated plants and ornamentals. The FAO estimates about $1.5 \times 10^9$ hectares of arable cropland; at a density of $10^5$--$10^6$ plants per hectare, there are $10^{13}$--$10^{14}$ plants in cultivation at any time. Assuming that 1--10\,\% are infected at any time yields the figure of $10^{12}$ plants infected used in the order-of-magnitude estimation.} (iii) the viroid load of any infected plant is $10^{12}$;\footnote{Although precise quantification of viroid copy numbers per plant is seldom reported, high-throughput sequencing and biochemical assays reveal that systemic viroid infections can produce extremely high intracellular titers. For large hosts such as fruit trees, the product of cell number and per-cell viroid load readily exceeds $10^{12}$ genomes, supporting the use of that scale in upper bound estimates.} and (iv) every replication of the genome includes on average one error (a new nucleotide). Assuming as well that all generated sequences are different (though many of them might be identical or very similar, and/or produce identical descent), we get an upper bound for the total number of explored sequences: $10^{36}$. Though large, this number is still far from covering the complete neutral network of the ASBVd secondary structure, not to mention a significant fraction of the whole space of constant $L$. Analogous figures for other genes and genomes yield values that make a systematic exploration of the set of sequences compatible with a phenotype absolutely out of reach \cite{louis:2016}. 

What do these numbers mean? In light of the high diversity and plasticity of life, and of the ability of organisms (notably microorganisms) to adapt to harsh and novel environments, it is clear that exploring most of sequence space is not a requisite for a population to thrive. As early studies with RNA postulated and illustrated numerically \cite{huynen:1996}, and more recent analyses suggest \cite{dryden:2008}, the specific identity of nucleotides (or amino acids) is not a requirement to attain function. 
%Local exploration (as opposed to the full exploration of the complete sequence space, or at least a whole neutral network) of variants must suffice.
Local exploration of variants---rather than exhaustive exploration of the full sequence space, or even of an entire neutral network---must suffice.

\section{Fitness landscapes, phenotype accessibility, and high-dimensional constraints}

A fitness landscape is a mapping from the genomic, multidimensional space to a real value representing fitness \cite{wright:1931}. Traditionally, fitness landscapes have been viewed as topographic representations where genotypes lie on the horizontal plane and fitness on the $z$-axis; this view has illustrated evolutionary dynamics for almost a century \cite{pigliucci:2012}. These landscapes portray the adaptive process as a parsimonious movement of populations that ``traverse valleys,'' ``reach fitness peaks,'' and adapt by incorporating mutations that pull them ``uphill.'' Inherent to these representations are concepts such as trapping in local maxima, peaks that can be attained, and, most of the time, fitness values with a sort of Platonic existence, independent of the state of the population. Although ample evidence shows that molecular evolution does not generally follow the iconic topographic fitness landscape, this metaphor remains deeply imprinted in our conceptual visualization of evolutionary dynamics.  

\subsection{Accessibility of fitness peaks in high dimensions}

Dynamics in multidimensional spaces differ fundamentally from two-dimensional projections. In high-dimensional genotype spaces, the accessibility of fitness peaks---that is, the existence of monotonic mutational paths leading from arbitrary genotypes to the global fitness maximum---depends critically on the ruggedness and correlation structure of the landscape. Analytical and numerical studies of random fitness models have shown that in uncorrelated or weakly correlated landscapes, such as the House-of-Cards or NK models with fixed local interactions, the probability that any strictly increasing path reaches the global optimum decays exponentially with sequence length $L$ \cite{franke:2011,krug:2019,schmiegelt:2023}. This loss of accessibility occurs because each beneficial step must satisfy a sequence of rare inequalities in fitness orderings, and in high dimensions these constraints multiply, making global fitness peaks effectively unreachable by purely uphill walks. Increasing the alphabet size $K$ (allowing more allelic states per site) or introducing a global additive gradient (as in the Mount Fuji model) can substantially increase the number of accessible trajectories, yet these conditions are atypical of natural, epistatic landscapes \cite{zagorski:2016,franke:2011}. 

While small empirical landscapes often exhibit multiple accessible routes \cite{szendro:2013}, the generic expectation for large genomes is that strictly monotonic adaptive paths are exceedingly rare. Reaching the global maximum typically requires transient declines in fitness or exploration of neutral or nearly neutral networks. Hence, the intuitive Wrightian picture of populations climbing continuous mountains is misleading in high-dimensional sequence spaces: despite the astronomical number of possible mutational paths, most are blocked by sign epistasis, and purely uphill trajectories almost never connect distant genotypes \cite{krug:2019}. 

\subsection{Invisible peaks and entropic barriers}

A more accurate description of evolutionary dynamics on sequence spaces should combine rough fitness landscapes with the partition of the space induced by mapping genotypes to phenotypes. Although genotype-to-fitness landscapes are often the default representation, considering the intermediate mapping to phenotypes illustrates an additional potential difficulty in reaching high-fitness solutions. The unified picture is that of adaptive multiscapes \cite{catalan:2017}, where each genotype maps to one phenotype (in a simplified many-to-one GP map), and all genotypes belonging to the same phenotype take the same fitness value in a specific environment. In this scenario, the highly skewed distribution of phenotype sizes comes into play: only the largest genotypes are relevant to evolution; fitter phenotypes, if sufficiently smaller, are almost never found. In other words, even if the global fitness peak (understood as a degenerate ``peak'' corresponding to all genotypes mapping to that phenotype) is accessible, it is likely lost in a sea of astronomically more abundant alternatives. 

\subsection{Phenotypic accessibility and the arrival of the frequent}

Mean-field models of GP maps provide a remarkably compact description of the high-dimensional dynamics of molecular evolution. In such models, the complex local structure of neutral networks is averaged out, and the transition probability $\phi_{\xi \to \chi}$ from phenotype $\xi$ to phenotype $\chi$ upon a point mutation can be computed as an average over all genotypes realizing $\xi$ \cite{schaper:2014,greenbury:2016,manrubia:2015,huynen:1996}. This approach predicts that, to a first approximation,   
\begin{equation}
\phi_{\xi \to \chi} \propto f_\chi \, ,
\end{equation}
where $f_\chi$ is the global frequency (or relative neutral set size) of phenotype $\chi$, i.e., the fraction of sequence space that maps to it. Because $f_\chi$ typically spans many orders of magnitude and is approximately log-normally distributed in RNA and other GP maps, the corresponding transition probabilities vary comparably. As a result, most of the mutational flux from any given phenotype leads to a very small subset of abundant, highly connected phenotypes, while rare phenotypes--—those with small size $N$---are practically inaccessible, even when they would confer large fitness advantages. This bias, termed the \emph{arrival of the frequent} \cite{schaper:2014}, implies that the spectrum of available variation is determined primarily by the statistical geometry of the GP map rather than by selective gradients on fitness. Note that the mean-field estimation is remarkably optimistic regarding transition probabilities: if there is a small number of links between the two phenotypes, and considering the huge sizes of the latter (even small ones), populations may evolve for extremely long times away from the apposition regions where the two phenotypes are one mutation apart. 

A simple numerical example illustrates the magnitude of this bias in the mean-field scenario. Consider sequences of length $L = 100$ with a four-letter alphabet (as in RNA), yielding $4^{100} \simeq 10^{60}$ possible genotypes. A typical large phenotype has size $N_\text{freq} \simeq 10^{42}$, the most abundant phenotype size is $10^{32}$, and typical small phenotypes have size $N_\text{rare} \simeq 10^{22}$ \cite{dingle:2015}. For a rare phenotype with a neutral set size $10^{20}$ times smaller, both theory and simulation suggest that, to a first approximation, the probability of a mutation leading into this phenotype scales roughly with its relative size $N$, so $\phi_{\text{rare}} / \phi_{\text{freq}} \simeq 10^{-20}$.\footnote{This ratio could be corrected by the fraction of neutral mutations for each phenotype size, but the correction is just a factor of 2, thus not modifying the order of magnitude.} Hence, even if a population explores $10^{12}$ {\em different} genotypes through mutation and drift, the expected number of encounters with that rare phenotype remains negligibly small. Conversely, neutral exploration within frequent phenotypes dominates: the population will perform an effectively {\em local} random walk within large, percolating neutral networks that connect vast regions of sequence space, even if large excursions from the initial sequence are uncommon. 

Analytical results allow estimating the ratio between the sizes of frequent and rare phenotypes as sequence length increases. For RNA, the distribution of the natural logarithm of phenotype sizes $\log N$ follows a normal distribution with mean $\mu_L= \mu_1L + O(1)$ and standard deviation $\sigma_L = \sigma_1 L^{1/2} + O(L^{-1/2})$, where $\mu_1=0.2865$ and $\sigma_1=0.4434$ \cite{garcia-martin:2018}. Then, the typical size of large phenotypes can be shown to be \cite{catalan:2023}
\begin{equation}
    \log N_{\text{freq}} \simeq \mu_L + \sigma_L^2 - z\sqrt{2}\sigma_L \, ,
\end{equation}
with $z \in (1.5,2.7)$ for the fraction of genotypes included in the set of frequent phenotypes, $p \in (0.99, 0.9999)$. By symmetry, since the distribution of phenotype sizes peaks at $\mu_L$, the typical size of rare phenotypes can be analogously estimated as 
\begin{equation}
    \log N_{\text{rare}} \simeq \mu_L - \sigma_L^2 + z \sqrt{2}\sigma_L \, .
\end{equation}
Hence, the mean-field estimate of the ratio of rare to frequent phenotype sizes is
\begin{equation}
\frac{\phi_{\text{rare}}}{\phi_{\text{freq}}} \simeq \exp \left(-2 \sigma_1^2 L + 2 \sqrt{2} z \sigma_1 \sqrt{L} \right) \, .
\end{equation}
This estimation yields larger relative values for $L=100$ than the rough numerical estimate obtained above, since it estimates smaller and larger phenotype sizes for frequent and rare phenotypes, respectively. However, the important result is that the ratio decreases exponentially with $L$, broadening the separation between the size of frequent and rare phenotypes and making rare phenotypes increasingly difficult to meet. For $L=246$, as in our viroid example, and taking $z=2$, $\phi_{\text{rare}} / \phi_{\text{freq}} \simeq 4 \times 10^{-34}$.

In summary, small, isolated phenotypes remain essentially invisible. Canalization or selective bias could, in principle, guide trajectories toward fitter phenotypes, but only within the subset of variation that the GP map makes accessible. This subset is essentially covered by other frequent phenotypes. 

Even if a population were artificially placed on one of these small, high-fitness phenotypes---analogous to fixing an adaptive RNA or protein structure in the laboratory---the phenotype would be rapidly lost under mutation pressure. Because most mutational trajectories leave the small neutral set, the population diffuses away in a few generations toward more abundant phenotypes with larger mutational robustness. This loss is fundamentally entropic: in the same way that systems in statistical physics evolve toward macrostates with more microstates, evolutionary dynamics naturally drift toward regions of sequence space with greater combinatorial weight. Thus, both the difficulty of locating rare phenotypes and the instability of remaining on them ensure that the ``arrival of the frequent'' dominates evolutionary exploration, limiting accessibility to the global fitness maximum (which, by probability, corresponds to the set of non-frequent phenotypes) except under extreme, non-generic conditions---an outcome consistent with results from models of fitness landscapes in high dimensions, where ruggedness and dimensionality conspire to render global peaks exponentially rare and typically unreachable \cite{franke:2011,nowak:2013}.

\begin{figure}
    \centering
    \includegraphics[width=0.7\linewidth]{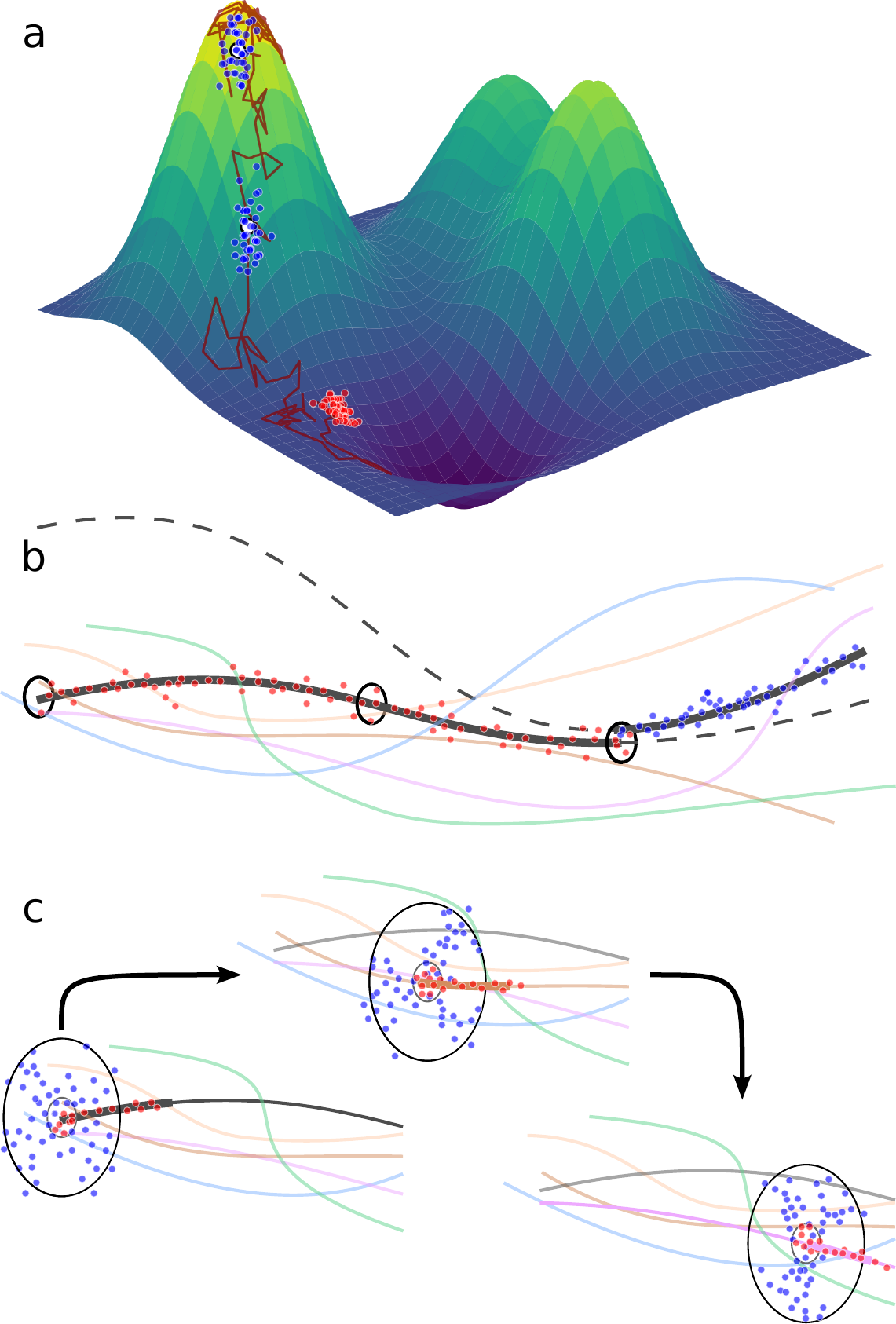}
\caption{{\bf Historical illustration of evolving viral quasispecies.} Our understanding of how viral quasispecies explore sequence space has evolved over time as empirical evidence has accumulated. 
%New perspectives do not discard earlier ideas, but rather retain their key insights while shifting attention toward features revealed by data.
({\bf a}) In early fitness‐landscape models, evolutionary dynamics were primarily described in terms of fitness increases. A simulated adaptive walk (brown curve with sampled red and blue points) illustrates this view. A viral population is assumed to start at low fitness, with individual variants independently sampling the local landscape. Fitter variants are selectively amplified, allowing the population as a whole to climb toward higher fitness peaks. Stochastic effects may occasionally enable escape from local optima. {\bf (b)} When most mutations are nonviable or incur large fitness losses, the simple fitness‐ascent picture becomes problematic. This motivated a shift in emphasis from fitness progression to neutral navigation. Neutral networks (colored curves) span genotype space, connecting widely different sequences that produce the same phenotype 
%(e.g., the same protein or RNA structure) 
through successive neutral mutations. A quasispecies can drift along such a network (thick black curve, red dots) without loss of fitness, while continually sampling nearby phenotypes. Occasionally, a previously distant neutral network (dashed black curve) comes into close proximity, offering access to higher fitness. The population then shifts to and explores this new network (blue dots). {\bf (c)} Recent empirical evidence suggests that viral quasispecies do not primarily rely on neutral network connectivity to navigate fitness landscapes. Instead, they generate a large fraction of all possible variants within a relatively broad mutational neighborhood (outer black circles) around the most abundant sequence, which is itself highly populated (blue dots). Neutral variants (red dots) are explored within the same radius, rather than preferentially. The schematic separation between neutral and non-neutral variation is shown for clarity; in reality, neutral networks are interwoven with non-neutral neighborhoods. Shifts in the dominant (reference) sequence occur frequently (twice in this illustration), after which the statistical structure of the explored neighborhood is rapidly re-established.
} 
\label{fig:newparadigm}
\end{figure}

\section{Efficient adaptation within high-abundance phenotypes}

The entropic, trapping effect caused by the large size of abundant phenotypes was described long ago \cite{schuster:1988} and observed in multiple numerical studies \cite{wolff:2009,khatri:2009,wilke:2001Nat,codoner:2006,cowperthwaite:2007,schaper:2014,catalan:2020}, where this effect manifests in various forms. Terms such as \textit{phenotypic entropy}, \textit{redundancy}, or \textit{landscape flatness} have been used to conceptually describe the number of genotypes that realize a given phenotype and, consequently, its mutational robustness. A pioneering theoretical analysis \cite{schuster:1988} considered the competition between two quasispecies corresponding to distinct phenotypes and showed that the phenotype with the highest replication rate could nevertheless be outcompeted by a slightly less fit one if the latter occupied a larger and more connected neutral network. Too low robustness, they demonstrated, could be fatal at high mutation rates. Note that in this analysis, phenotype size was not included as a feature of fitness, and only the replicative ability of the genotypes within the phenotype was considered. An analytical treatment \cite{wolff:2009} later quantified the conditions for this crossover between fitness and robustness, showing that increasing mutation rates can drive the dominance of phenotypes with lower intrinsic fitness but broader mutational neighborhoods, and that this transition is closely related to the onset of the error catastrophe. 

Other studies have examined analogous effects in different biological and computational contexts. In models of spatial gene regulation, it was shown that finite populations often fail to converge to the globally optimal phenotype when multiple equivalent solutions exist, an outcome interpreted as entropic and arising from the multiplicity of equally viable configurations \cite{khatri:2009}. The same principle underlies the so-called ``survival of the flattest'' phenomenon \cite{wilke:2001Nat}, where, at sufficiently high mutation rates, populations evolve toward flatter regions of the fitness landscape, prioritizing mutational robustness over maximal replication speed. This prediction was later confirmed experimentally with RNA viroids \cite{codoner:2006}, where lineages displaying slower replication but higher neutrality outcompeted faster yet fragile variants, showing that fitness can be maximized by minimizing the mutational load rather than by maximizing growth rate. 

Computational analyses of mutation–selection dynamics have further clarified the evolutionary implications of phenotype size. It has been shown that adaptation tends to proceed through abundant phenotypes---a process also termed the ``ascent of the abundant'' \cite{cowperthwaite:2007}---and that phenotypic innovation is biased toward more frequent phenotypes in genotype–phenotype maps, the effect already described as the ``arrival of the frequent'' \cite{schaper:2014}. More recent analytical work \cite{catalan:2020} has demonstrated that this bias persists across a broad range of mutation–selection regimes, confirming that phenotypic accessibility, rather than differences in maximal fitness, largely determines the evolutionary trajectories available to finite populations.

In summary, only a vanishingly small fraction of highly abundant phenotypes is explored along evolution and thus presented to natural selection. The question of whether this restriction limits, and to what extent, the localization of phenotypes with sufficiently high fitness was addressed in a study where the relative weight of phenotype size versus replicative ability was analytically and numerically explored \cite{catalan:2023}. Returning to the case of RNA sequence-to-secondary structure GP maps, the fraction $u$ of phenotypes visible to evolution diminishes exponentially with $L$ as
\begin{equation}
    u \simeq \frac{0.9}{L^{1/2}}(1.1)^{-L} \, .
\end{equation}
However, since the total number of different open secondary structures for sequences of length $L$ grows as $1.48 L^{-3/2}(1.85)^L$, the absolute number of different, large phenotypes covered by a fraction $p\simeq 1$ of genotypes grows exponentially with $L$ as $1.33 L^{-2}(1.68)^L$. 

Assuming, as in \cite{catalan:2023}, that replicative ability values follow a normal distribution with average $\langle r \rangle$ and variance $\sigma_r^2$, even in the restricted set of large phenotypes visible to natural selection, the maximum value of $r$ that can be attained is only about 10\,\% lower than in the whole ensemble of phenotypes. In other words, limiting the evolutionary search to large phenotypes does not entail a large fitness loss, while it can yield a significant gain in robustness and navigability. 

The exploration of phenotypic diversity is not only restricted to large phenotypes, but it is also local in sequence space. No better demonstration of the locality of evolution exists than gene phylogenies, where common ancestors occurred billions of years ago. Retaining memory of this distant past indicates that extant, successful functions were not far from the sequences where the search started. This observation relates to the idea of shape space covering discussed above, which appears to be a general feature of the broad sequence-to-function map: new functions may lie a few mutations away from current genotypes in high-dimensional spaces. Support for this expectation, originally derived from simple GP maps, has come from experimental results showing that functional genes can be rapidly generated from non-genic sequences \cite{carvunis:2012,vakirlis:2020}, that transposable elements can be domesticated to perform specific functions in their host \cite{sinzelle:2009}, or that promoters can emerge from random sequences \cite{yona:2018}, even in the absence of sequence diversity, simply through successive cycles of mutation, enrichment, and selection \cite{wachowius:2021}. Thus, high-dimensional sequence spaces allow the discovery and fixation of molecular function diversity observed in current organisms, speeding up adaptive rates so much as to allow organisms to thrive in steadily changing environments.

\section{Hierarchical genotype networks in empirical populations}

A conceptual metaphor that helps us understand how evolutionary dynamics unfold in sequence spaces needs to integrate the main elements discussed in previous sections: (i) sequence spaces are so vast that only a minute fraction is explored by natural populations; (ii) neutral networks are astronomically large, and natural populations only visit a local, tiny fraction of the whole; (iii) spaces of high dimensionality are so deeply interwoven that most (large) phenotypes can be found a few mutations away; (iv) as sequence length increases, high-fitness maxima become increasingly difficult to attain; (v) the fraction of (large) phenotypes explored by genotypes of length $L$ decreases exponentially with $L$, though at the same time, the absolute number of different phenotypes increases exponentially; (vi) the latter restriction has a small impact on the replicative ability values that populations can achieve. Establishing a mindset within this scenario should also be helpful when interpreting empirical results, and should prove especially relevant when heterogeneity is intrinsic to the population, such as in viruses and other organisms with high genotypic diversity. These populations should comply with the restrictions above while managing to explore sufficient evolutionary innovation or variants that allow their adaptation to changing conditions. A pictorial summary of how the elements above modify suitable representations of quasispecies evolution on sequence spaces is represented in Fig. \ref{fig:newparadigm}. 

Advances in sequencing technologies have boosted the detailed description of genetic diversity in viral populations in particular. This has allowed the characterization, to varying but deep detail, of the within-host structure of viral populations \cite{baaijens:2017,zanini:2017,delgado:2021} and their phylodynamics \cite{attwood:2022,sun:2025}. Recently, deep sequencing of the Q$\beta$ bacteriophage has revealed a hierarchical structure with implications for viral evolutionary dynamics and, in particular, for the search process in sequence spaces \cite{seoane:2025}. 

\begin{figure}
    \centering
    \includegraphics[width=\linewidth]{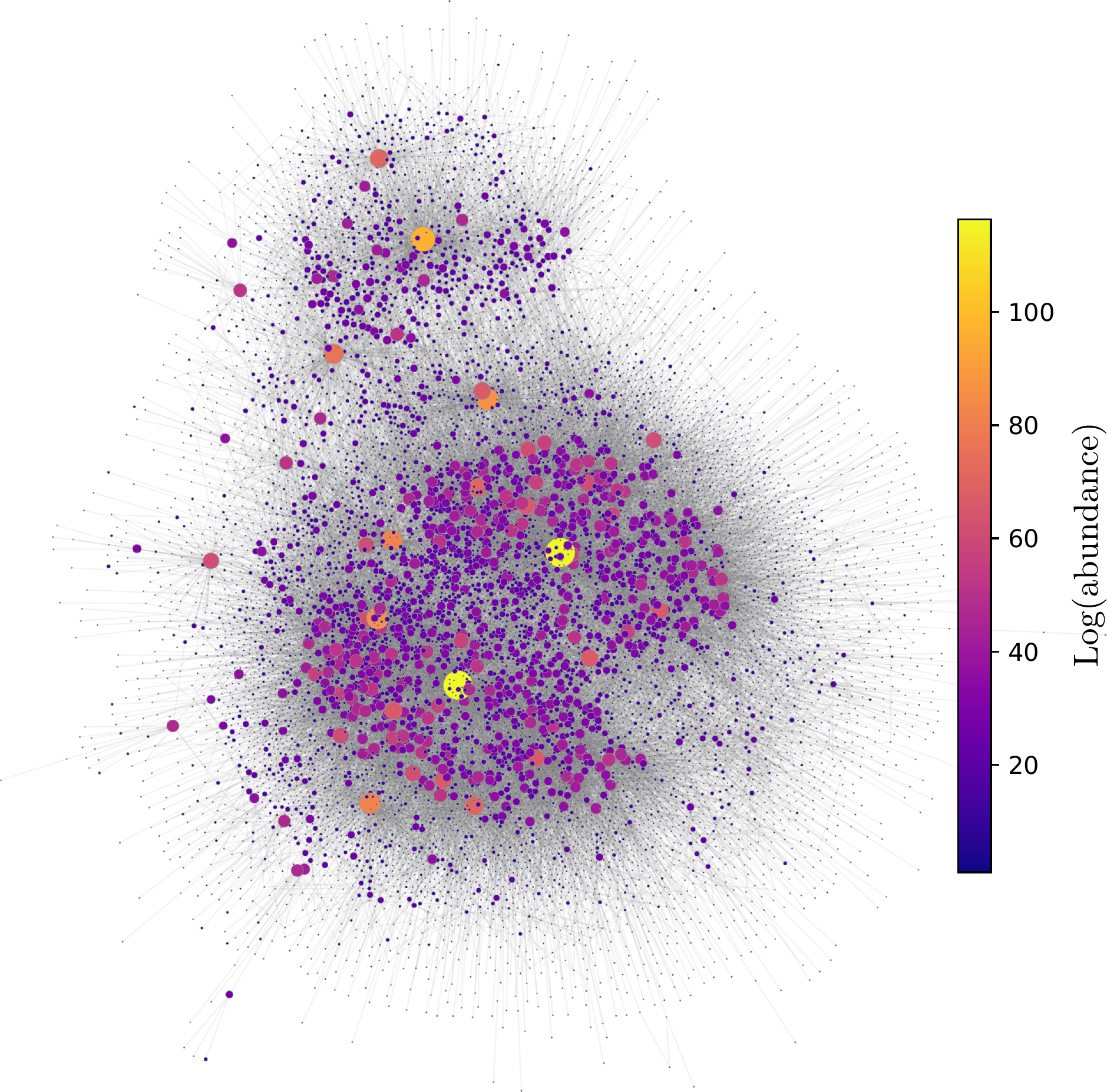}
\caption{{\bf An empirical genotype network.} 
%We performed an evolution experiment on the Q$\beta$ phage (details in \ref{XXX}). We used deep-sequencing techniques to sample the viral quasispecies at the end of the experiment (we work with passage $60$ at $43\circ$C from \ref{XXX}). We sorted and aligned all sequences obtained, and retained those of a same length ($271$ nucleotides). In the plots, 
Each node in this network represents a unique sequence (haplotype) of a Q$\beta$ phage population; two haplotypes are connected if they differ in just one nucleotide. Data correspond to passage 60 of a population adapting to a temperature of $43^{\circ}$C. A full description of the deep sequencing data and the curation protocol can be found in \cite{seoane:2025}. Node size is proportional to the logarithm of the abundance of the haplotype (the number of sequences identical to that haplotype), and color conveys the same information. The spring layout (which positions nodes by treating them as masses connected by springs and minimizing tensions) suggests community structure around a series of salient nodes. These seem to act as {\em sources} of viral replicas which, through mutation, populate a neighborhood around each network hub. The self-similar structure of the network is reflected, among others, in a robust power-law degree distribution \cite{seoane:2025}.
%These mutants are likely less fit than the hubs, thus take away from the global replicative dynamics of the quasispecies---in essence behaving as {\em sinks} of the flow from the hubs towards their periphery. {\bf b} A radial representation with the root (most abundant node) at center further reveals a self-similar, hierarchical structure in the suggested {\em source-to-sink} dynamics. Somehow fit mutants at each fixed distance from the root re-enact the source motif spreading the outflux further away from the network center. Power-law distributions of several network properties strongly support the fractal self-organizing nature of viral quasispecies.
}
\label{fig:QbetaNet}
\end{figure}

Q$\beta$ populations exhibit robust organizational patterns along evolution, even when new mutations with a supposed adaptive value become fixed. Figure \ref{fig:QbetaNet} represents a snapshot of a genotype network of a deep-sequenced Q$\beta$ population. At any time point in a process of adaptation to new environments, genomes cluster around the most abundant sequence, which appears to continuously regenerate diversity, acting as a diversity source and populating its neighborhood at increasing mutational distance in a process dominated by simple diffusion: sequence abundance decays exponentially with Hamming distance to the most abundant sequence. A remarkable observation is that, on average, no significant correlation between the abundance of variants and the type of mutation they carry (synonymous or non-synonymous) is found. In this out-of-equilibrium state, the search process is dominated by reaction (the most abundant sequence produces a larger number of offspring) and diffusion (through mutation), with a minor effect of the underlying fitness landscape. As new mutations fix in the population, the whole hierarchical structure is rapidly reconstructed around the new center. Figure \ref{fig:QbetaNet} illustrates the uncommon instance when the population transitions between two dominant haplotypes. 

These observations illustrate how heterogeneous populations explore sequence space and, with it, new potential phenotypes: little use of neutral paths seems to occur, with sequential displacements being apparently more dependent on mutations with at least some adaptive advantage. At the same time, almost all possible variants one- and two-mutations away, and a significant fraction at larger distances, are generated by a massive production of mutants before purifying selection can eliminate low-fitness variants. Therefore, fitness ``valleys'' no longer confine the population to high-fitness variants. In this scenario, the concept of a fitness peak becomes diffuse: population dynamics only see a coarse representation of the landscape, where the fitness of individual sequences is no longer the unique determinant of their fate---whether maintained or discarded. This provides an alternative view of quasispecies as units of selection, since their structure and dynamics are in mutual feedback to guarantee persistence.

\section{Conclusions}

Evolution and adaptation occur primarily at a local level in sequence space. The astronomical size of sequence spaces, even of the subsets of genotypes mapping to the same phenotype, strongly suggests that this cannot be otherwise. Evidence for limited displacements in these vast spaces since functional solutions were first located is provided by gene phylogenies that can be traced hundreds of millions of years back in time: long excursions through neutral drift are not essential for adaptation.

Despite this local exploration, the high dimensionality of sequence spaces causes abundant phenotypes to be highly interwoven, with regions of contact that are frequent and occur in many separated regions of sequence space. Consequently, a limited exploration of genotype space does not imply a limited exploration of phenotype space. This conclusion extends to the exploration of fitness space, assuming that the distribution of fitness values has a well-defined average \cite{catalan:2023}---an assumption supported by the variations in functionality observed in natural systems.

A particularly striking result is that Maynard Smith's notion of a connected protein space, which allows the sustained accumulation of mutations without substantial fitness loss, is not required for the evolution and adaptation of viral populations. Instead, it suffices that phenotype abundances be highly biased, ensuring that frequent phenotypes lie close to one another in sequence space. Viral populations generate and explore all available diversity at short mutational distances from fit, abundant sequences, largely independently of the detailed structure of the underlying fitness landscape. This finding is relevant for understanding the emergence of function in populations of replicators, both within cellular environments (as may have occurred during the emergence of viroids \cite{catalan:2019a}) and in pre-cellular evolutionary contexts \cite{manrubia:2022}.

We may even entertain the idea that percolation of sequence space by neutral networks, and the notion that it facilitates navigation of genotype spaces, are merely spandrels of the properties of large phenotypes that are actually relevant for evolution: they exist due to their abundance and persist due to entropic principles. Only a minute fraction of neutral networks is ever visited in evolutionary time, a fact that does not contradict the emergence of molecular function through repeated selection cycles on random sequences: functional sequences are often just a few mutations away.

\begin{acknowledgement}
The authors acknowledge the contributions of Jacobo Aguirre, Iker Atienza, Pablo Catal\'an, Juan Antonio Garc\'{\i}a-Mart\'{\i}n, Ester L\'azaro, Samuel Mart\'{\i}nez-Alcal\'a, Henry Secaira-Morocho, and Ariadna Villanueva to the development of the unified picture presented here. This work was funded by MICIU/AEI/10.13039/501100011033 and by ERDF/EU (PGE) `A way of making Europe', through grants PID2023-147963NB-C21 (SM), PID2023-153225NA-I00 (LFS) and PID2022-141802NB-I00 (BASIC) (JAC). 

\end{acknowledgement}
%
%\section*{Appendix}
%\addcontentsline{toc}{section}{Appendix}
%
%

%\input{references}

\bibliographystyle{plain}
\bibliography{bibliography}

\end{document}